\documentclass[preprintnumbers, twocolumn, amsmath, amssymb, aps, prb, 10pt, showkeys]{revtex4-2}

\usepackage{graphicx}
\usepackage{xcolor}
\usepackage{dcolumn}
\usepackage{bm}
\usepackage{braket}
\usepackage{siunitx}
\usepackage[bookmarks=false,
            breaklinks=true,
            colorlinks=true,
            linkcolor=blue, 
            urlcolor=blue, 
            citecolor=blue]{hyperref}
\begin{document}

\preprint{}

\title{Influence of the Effective Mass on \textit{ab initio} Phonon-limited Electron Mobility of GaAs}

\author{Mohammad Dehghani$^{1,2}$}
\email{dehghani@iue.tuwien.ac.at}
\thanks{\href{https://orcid.org/0000-0003-4756-0446}{ORCID: 0000-0003-4756-0446}}

\author{Dominic Waldhoer$^{1,2}$}
\email{waldhoer@iue.tuwien.ac.at}
\thanks{\href{https://orcid.org/0000-0002-8631-5681}{ORCID: 0000-0002-8631-5681}}

\author{Angus Gentles$^{1,3}$}

\author{Pedram Khakbaz$^{1}$}

\author{Rainer Minixhofer$^{3}$}

\author{Michael Waltl$^{1,2}$}

\affiliation{$^1$Institute for Microelectronics, TU Wien, Gußhausstraße 27-29, Wien, A-1040, Austria}
\affiliation{$^2$Christian Doppler Laboratory for Single-Defect Spectroscopy at the
    Institute for Microelectronics, TU Wien, Vienna, Austria}
\affiliation{$^3$ams-OSRAM AG, Tobelbader Str. 30, 8141 Premst{\"a}tten, Austria}

\date{\today}

\begin{abstract}
We present a comprehensive \textit{ab initio} study of the influence of band structure corrections, particularly the electron effective mass, on the phonon-limited electron drift and Hall mobilities of GaAs. Our approach is based on the  DFT+$U$ method, combined with an iterative solution of the linearized Boltzmann transport equation using the Wannier interpolation technique. We show how this framework allows for accurate refinements of the electronic band structure and phonon dispersion, leading to improved predictions for transport properties. In particular, by varying the Hubbard parameters to purposefully tune the conduction band features, allowing us to reproduce bands with different electron effective mass, we systematically investigate the relationship between mobility and effective mass. In this context, our results show close agreement with semi-empirical relations that follow a power-law dependence. Moreover, this approach can be used to indirectly incorporate temperature effects into the band structure, enabling efficient evaluation of temperature-dependent electron mobilities. Our mobility results exhibit good agreement with experimental data and are comparable to previously reported values obtained using the computationally expensive GW method.
\end{abstract}


\maketitle

\section{\label{sec:intro} Introduction}
Carrier mobilities are a crucial property affecting the performance of semiconductor devices. As semiconductor technologies advance and increasingly complex functional materials emerge, predicting and optimizing carrier mobilities through simulations is essential to meet the evolving demands of modern electronics. Over the last decade, \textit{ab initio} methods based on density functional theory (DFT) have been developed to predict carrier mobilities, providing invaluable atomistic-level insights to electronic transport~\cite{EPW_code_2023, Perturbo_code_2021, AO_bernardi_2018, Abinit_mobility_fourier_brunin}.

\par In all these methods, the linearized Boltzmann transport equation (LBTE)  is solved for the carrier distributions in the presence of electric fields and electron-phonon (e-ph) scattering. The central quantity in this framework is the e-ph coupling matrix, $g_{mn}^{\nu}(\mathbf{k}, \mathbf{q})$, whose elements describes the coupling between two Bloch states $\ket{m\mathbf{k}}$ and $\ket{n\mathbf{k}'}$,  mediated by
a phonon of branch \(\nu\) and wavevector
$\mathbf{q} = \mathbf{k}' - \mathbf{k}$. Achieving numerically converged results requires evaluating this matrix on dense $\mathbf{k}$- and $\mathbf{q}$-grids. Although density functional perturbation theory (DFPT)~\cite{DFPT_Baroni2001} can, in principle, be used to compute these elements, its application becomes impractical due to the high computational cost associated with such fine grids. To overcome this limitation, the matrix elements are typically interpolated onto dense grids using DFT and DFPT data calculated on coarser meshes. The primary distinction among available methods lies in the interpolation scheme~\cite{nature_Rev_phys_Claes2025}, with the Wannier function (WF)-based approach~\cite{Giustino2007, Calandra_2010} being the most adopted method in recent years. Several widely used software packages, including \textsc{EPW}~\cite{EPW_code_2023} and \textsc{Perturbo}~\cite{Perturbo_code_2021}, have been developed based on this technique and are extensively applied in the study of carrier transport in solid-state materials.

\par Within these packages, several methods are available to compute the carrier mobility by solving the LBTE. Historically, the first approach is the relaxation time approximation (RTA), which assumes that the relaxation time equals the lifetime of a charged quasiparticle excitation. This lifetime is proportional to the inverse of the imaginary part of the e-ph Fan-Migdal self-energy \cite{Giustino2007}. This method is often referred to as the self-energy RTA (SERTA), where the total scattering rate then follows by integration over the Brillouin zone. However, the SERTA systematically underestimates carrier mobilities due to its inability to distinguish between backward and forward scattering processes, treating all scattering events as equally impactful regardless of changes in carrier velocity \cite{epw_ponce_Hall_mobility_2021}. To address this limitation, the momentum relaxation time approximation (MRTA) has been introduced, in which scattering rates are weighted by relative changes in carrier velocities during each scattering event \cite{ponce_review_iop_2020}, improving mobility estimates by better capturing the actual momentum relaxation process. However, despite being more accurate than SERTA, MRTA still results in poor mobility prediction for many materials~\cite{assessing_rta_Claes2022}. A more accurate but also computationally more demanding method is the iterative approach (ITA) \cite{ITA_Fiorentini2016, ITA_Li2015, Perturbo_code_2021}. This method provides an exact solution to the LBTE and has been shown to significantly improve mobility values compared to both SERTA and MRTA \cite{ponce_review_iop_2020, assessing_rta_Claes2022, epw_ponce_Hall_mobility_2021}. 

\par A major challenge in obtaining reliable mobility values in semiconductors is that standard exchange-correlation functionals in DFT, such as the local density approximation (LDA) and the generalized gradient approximation (GGA), often fail to accurately describe the band structure of many materials \cite{narrowbandgap1, narrowbandgap2, narrowbandgap3}. Since carrier mobility is highly sensitive to the details of the band structure, especially near band edges, these inaccuracies can strongly influence transport results. In reported \textit{ab initio} transport studies, a widely adopted solution is the GW method, which fundamentally corrects the electronic band structure by accounting for many-body interactions through the screened Coulomb potential, providing more accurate quasiparticle energies compared to DFT. Numerous studies have demonstrated improvements in predicted transport properties using this method in combination with DFPT and ITA \cite{GaAs_GW_EPW_Liu2017, GaAs_EPW_Ma2018, EPW+GW_diamond_Shoemaker2024, ponce_epw+GW_PRB2018}. However, despite its effectiveness, the high computational costs of the GW method can be a major hindrance to its use for complex systems.

\par Polar semiconductors constitute an important class of materials widely used in optoelectronics and for power devices. In particular, gallium arsenide (GaAs) is a common benchmark material for which carrier mobilities have been extensively investigated using \textit{ab initio} methods~\cite{Perturbo_code_2021, GaAs_SERTA_Zhou2016, perturbo_GaAs_electron-two-phonon_Lee2020, GaAs_GW_EPW_Liu2017, GaAs_EPW_Ma2018, Abinit_mobility_fourier_brunin, quadrupole_GaAs_Brunin2020, epw_ponce_Hall_mobility_2021}. A significant achievement in these studies has been the development of efficient techniques to handle the divergent behavior of e-ph matrix elements near the $\Gamma$ point, which is an inherent characteristic of polar interactions~\cite{GaAs_SERTA_Zhou2016, frohlich_effect_Verdi2015, quadrupole_GaAs_Brunin2020}. Most of these studies rely on LDA for calculating the electronic structure and the phonon properties, as LDA yields reasonably accurate phonon band structures within DFPT~\cite{GaAs_SERTA_Zhou2016, GaAs_EPW_Ma2018}. When spin-orbit coupling is included, LDA also reproduces sufficiently accurate valence band characteristics, leading to satisfactory hole mobility results using ITA~\cite{GaAs_EPW_Ma2018}. However, a notable issue in these studies is that SERTA results for the electron mobility appear to be in good agreement with experimental values for GaAs, while ITA results overestimate the mobilty, an outcome attributed to LDA's underestimation of the electron effective mass~\cite{GaAs_EPW_Ma2018, Perturbo_code_2021}. However, combining GW-refined band structures with DFPT results from LDA resolves this discrepancy~\cite{GaAs_GW_EPW_Liu2017, GaAs_EPW_Ma2018}, yielding ITA mobility values aligned with experimental data and, as expected, underestimated SERTA results, which highlights the sensitivity of mobility predictions to the electron effective mass.

\par The sensitivity of carrier mobilities to the details of the band structure is a well-established fact~\cite{Lundstrom2000}. In the case of electron mobilities in polar semiconductors like GaAs, previous studies based on solving BTE using the variational technique~\cite{Ehrenreich1960, Ehrenreich1961}, Rode's iterative scheme~\cite{Rode1975} and Monte Carlo approach~\cite{MC-GaAs-Fischetti1988} confirm that this sensitivity is particularly tied to the electron effective mass and the nonparabolicity of the conduction band. Moreover, semi-empirical models, derived analytically for polar-optical-phonon scattering, exhibit an explicit dependence of the electron mobility~\cite{Ehrenreich1960, analytical_relation_polar_mobility_Gelmont1995} and the e-ph scattering rates~\cite{Lundstrom2000, GaAs_GW_EPW_Liu2017} on the effective mass. This, together with the fact that \textit{ab initio} methods, while preserving the nonparabolic character of the band, yield varying values for the effective mass, motivates a systematic investigation of the relationship between electron mobility and effective mass in \textit{ab initio} simulations, an investigation that, to the best of our knowledge, has not yet been carried out and is the focus of this work.

\par A potential approach to addressing the limitations of LDA and GGA in accurately describing electronic properties in insulators and semiconductors is the DFT+$U$ method \cite{DFT+U_Anisimov1991,DFT+U_Anisimov1996,DFT+U_Anisimov1997,DFT+U_Dudarev, DFT+U_Liechtenstein, DFT+U+J}. It introduces a self-energy correction term based on the Hubbard model, applied selectively to specific electronic states via projection operators, while treating the remaining states with standard LDA or GGA. Originally developed for correlated electron systems, DFT+$U$ improves the description of on-site Coulomb interactions of localized electrons in partially filled $\mathrm{d}$ and $\mathrm{f}$ orbitals, which are inaccurately described by LDA and GGA. Among different flavors of DFT+$U$, Dudarev's formulation \cite{DFT+U_Dudarev} is widely used, as it considers only the effective Hubbard parameter $U_\mathrm{eff} = U-J$, where $U$ and $J$ are on-site Coulomb and exchange Hubbard parameters, respectively. This method also has been extended to DFPT+$U$ \cite{DFPT+U_1, DFPT+U_2} for lattice dynamics calculations and successfully applied to phonon structure evaluations. Additionally, it has been employed to study e-ph interactions in CoO \cite{DFT+U_Perturbo} and transport properties of SrVO$_3$ \cite{Perturbo_DFT+U_Abramovitch2024}. 

\par Beyond its traditional applications in correlated electron systems, DFT+$U$ has also been successfully applied to semiconductors with partially filled $\mathrm{p}$ orbitals, demonstrating improved band structure predictions~\cite{DFTU_Kolesov2019, Bayesian_DFT+U_Popov, DFT+U_InN_Terentjevs2010, DFT+U_InN_sanchez2010, Bayesian_DFT+U_Yu}. Although \textit{ab initio} computational frameworks exist for determining Hubbard parameters from first principles, many studies instead employ machine learning techniques, particularly Bayesian optimization, to determine suitable $U$ values~\cite{Bayesian_DFT+U_Tavadze, Bayesian_DFT+U_Yu, Bayesian_DFT+U_Popov}. This approach adjusts the system's band structure as a function of Hubbard parameters to best match a reference band structure obtained either from experiment or from higher-level computational methods such as GW or hybrid functionals.

\par In this work, the phonon-limited drift and Hall electron mobilities of GaAs are investigated by solving the LBTE within ITA as implemented in the \textsc{Perturbo} package. To improve the accuracy of the band structure and phonon properties, DFT+$U$ and DFPT+$U$ methods are applied. Our study focuses exclusively on electrons, as accurately modeling holes in GaAs requires incorporating spin-orbit coupling to capture valence band splitting, which is essential for reliable hole mobility values~\cite{GaAs_GW_EPW_Liu2017}. However, non-collinear calculations, necessary for this purpose, are currently unsupported in this framework and require further methodological development. In contrast, such considerations are unnecessary for the conduction band, as no significant band splitting occurs within the transport energy window. First, we demonstrate how this approach improves the band structure and yields accurate phonon properties for this system. Then, we show that convergence of the mobility values with respect to the coarse $\mathbf{k}$ grid is tightly linked to the convergence of the effective mass value for the Wannier-interpolated bands. Furthermore, we show how this approach enables refinement of the conduction band features, particularly the effective mass and the $\Gamma$-L valley, allowing to explicitly study the electron mobility as a function of the effective mass. Finally, we compare our results with both experimental data and previous computational studies.

\section{\label{sec:results}Results and discussions}
\subsection{\label{subsec:band-structure}Electronic Structure and Phonon Properties}
Our DFT calculations are performed using \textsc{Quantum Espresso}~\cite{QE1, QE2, QE3} with scalar relativistic standard precision pseudopotentials~\cite{Pseudo-Dojo} based on the GGA-PBE~\cite{PBE} exchange correlation functional, where the semicore $3d$ states of both Ga and As atoms are explicitly treated and included in the valence configuration of the pseudopotentials. To improve the band structure and phonon properties calculated by the GGA-PBE approach, we apply the Dudarev formulation of DFT+$U$~\cite{DFT+U_Dudarev} and its linear response variant, DFPT+$U$~\cite{DFPT+U_1, DFPT+U_2}, incorporating Hubbard corrections on the $4p$ orbital of both Ga and As atoms. Hereafter, we refer to $U_{\mathrm{eff}}$ as $U$ for simplicity and DFT+$U$ as PBE+$U$ for clarity. Band structure calculations are carried out using a cutoff energy of the wave functions of 110 Ry, and a $12\times12\times12$ $\Gamma$-centered Monkhorst-pack $\mathbf{k}$-mesh. To calculate the phonon structure and dynamical matrices using DFPT, these parameters have been increased to 150 Ry and a $20\times20\times20$ grid on an $8\times8\times8$ $\mathbf{q}$-point mesh, to ensure well-convergent results. 

\par The Hubbard parameters, $U_{\mathrm{Ga}}^{4p}$ and $U_{\mathrm{As}}^{4p}$, are determined using a variant~\cite{Angus_syspad} of Bayesian optimization introduced by \citeauthor{Bayesian_DFT+U_Yu}. In this formalism, the loss function is defined as the root mean square difference between the calculated band structure and the reference at the high-symmetry points $\Gamma$, $X$, and $L$, for a given set of Hubbard parameters. Since the primary objective in Ref.~\cite{Angus_syspad} was to obtain an accurate direct band gap at $\Gamma$, a weighting factor of 0.75 is applied to the band difference at $\Gamma$, while factors of 0.25 are assigned to $X$ and $L$, thereby prioritizing the band gap accuracy over the valley positions. The loss function is then minimized through a Bayesian process by adjusting the Hubbard parameters to achieve an optimal band structure, in agreement with the reference data. 

\par In this work, we used the band structure obtained using HSE06 hybrid functional with default exchange fraction ($\alpha=0.25$) as the reference. This choice has been proved within previous studies~\cite{Bayesian_DFT+U_Yu, Bayesian_DFT+U_Popov} to yield reliable results for various materials, including semiconductors such as InAs and InGaN superlattices. To this end, the HSE06 band structure was computed using the same computational settings as described earlier. The resulting band structure was then used to find the optimized $U$ parameters as outlined above. The final values obtained were $U_{\mathrm{Ga}}^{4p}=~$\SI{0.16}{\electronvolt} and $U_{\mathrm{As}}^{4p}=~$\SI{3.50}{\electronvolt}. These parameters were subsequently employed within DFPT+$U$ calculations to compute the phonon dispersion and dielectric constants, which were compared against available experimental data.

\par Table~\ref{tab:table1} presents the equilibrium lattice constant and band structure parameters calculated using the PBE+$U$ method with the optimized Hubbard parameters mentioned above, alongside results obtained using HSE06 and GGA-PBE functionals. 
\begin{table}[htbp]
\caption{\label{tab:table1} Comparison of material parameters for GaAs calculated using PBE+$U$, HSE06, and PBE methods with available experimental data. Shown parameters: lattice parameter $a_0$; band gap $\Delta E_g$; the energies of the $L$ and $X$ valleys in the conduction and valence band ($\Delta E_{L-\Gamma}^{c}$, $\Delta E_{X-\Gamma}^{c}$, $\Delta E_{L-\Gamma}^{v}$ and $\Delta E_{X-\Gamma}^{v}$) with respect to the corresponding CBM and VBM; electron ($m^*_e$) and hole $(m^*_h)$ effective masses in units of the electron mass \unit{m_0}; dielectric constants $\varepsilon_0$ and $\varepsilon_{\infty}$; and optical phonon frequencies at the $\Gamma$.} 
\begin{ruledtabular}
\begin{tabular}{lcccr}
&\textrm{PBE+$U$} &\textrm{HSE06} &\textrm{PBE} &\textrm{Exp} \textrm{[Ref.]}\\
\colrule
\textrm{$a_0$(\AA)}                     &5.621 &5.631 &5.749 &5.642 \cite{madelung-2013}\\
\textrm{$\Delta E_g$ (eV)}              &1.36  &1.32  &0.15  &1.52 \cite{madelung-2013}\\
\textrm{$\Delta E_{L-\Gamma}^{c}$ (eV)} &0.20  &0.46  &0.72  &0.30 \cite{cband_minima_aspnes}\\
\textrm{$\Delta E_{X-\Gamma}^{c}$ (eV)} &0.71  &0.91  &1.27  &0.46 \cite{cband_minima_aspnes}\\
\textrm{$\Delta E_{L-\Gamma}^{v}$ (eV)} &-1.25 &-1.29 &-1.21 &-1.30 \cite{vband_maxima_chiang}\\
\textrm{$\Delta E_{X-\Gamma}^{v}$ (eV)} &-3.06 &-2.99 &-2.65 &-2.80 \cite{vband_maxima_chiang}\\
\textrm{$m^*_e$}                        &0.068 &0.061 &0.009 &0.067 \cite{experimental_effective_mass_Molenkamp}\\
\textrm{$m^{*[100]}_{\mathrm{lh}}$}              &0.066 &0.061 &0.009 &0.094 \cite{experimental_effective_mass_Molenkamp}\\
\textrm{$m^{*[110]}_{\mathrm{lh}}$}              &0.081 &-     &0.010 &0.081 \cite{vurgaftman2001}\\
\textrm{$m^{*[111]}_{\mathrm{lh}}$}              &0.054 &-     &0.011 &0.082 \cite{madelung-2013}\\
\textrm{$m^{*[100]}_{\mathrm{hh}}$}              &0.334 &-     &0.366 &0.34 \cite{experimental_effective_mass_Molenkamp}\\
\textrm{$m^{*[110]}_{\mathrm{hh}}$}              &0.334 &-     &0.367 &0.64 \cite{vurgaftman2001}\\
\textrm{$m^{*[111]}_{\mathrm{hh}}$}              &0.783 &-     &0.853 &0.75 \cite{madelung-2013}\\
\textrm{$\varepsilon_{0}$}              &12.24 &-     &54.75 &12.44 \cite{Moore1996}\\
\textrm{$\varepsilon_{\infty}$}         &10.21 &-     &52.54 &10.57 \cite{Moore1996}\\
\textrm{$\omega^{\mathrm{TO}}_{\Gamma}$ (THz)}      &7.93  &-     &7.56  &8.13 \cite{Strauch_1990}\\
\textrm{$\omega^{\mathrm{LO}}_{\Gamma}$ (THz)}      &8.86  &-     &7.72  &8.79 \cite{Strauch_1990}\\
\end{tabular}
\end{ruledtabular}
\end{table}
Additionally, it includes dielectric constants, and phonon frequencies at the $\Gamma$ point, calculated within DFPT and DFPT+$U$ methods. Here, the energy differences at high-symmetric points, $L$ and $X$, for both conduction and valence bands are referenced to the conduction band minimum (CBM) and valence band maximum (VBM), respectively. The effective masses were determined using the finite difference method, limited to points within 0.1-1 meV of the corresponding valley maximum or minimum. As shown in this table, PBE+$U$ and HSE06 yield comparable results for most parameters, with the notable exception of $\Delta E_{L-\Gamma}^{c}$, where HSE06 overestimates the value by more than 50\% relative to the experimental value, while PBE+$U$ underestimates it by 33\%. This discrepancy may stem from different weighting factors applied during optimization to prioritize the band gap over valley positions. The substantial improvement of PBE+$U$ results over PBE is evident from the tabulated data, particularly, in the accuracy of effective masses, dielectric constants, and phonon frequencies at the $\Gamma$ point.

\par
\begin{figure}[htbp]
\centering
\includegraphics[scale=0.42]{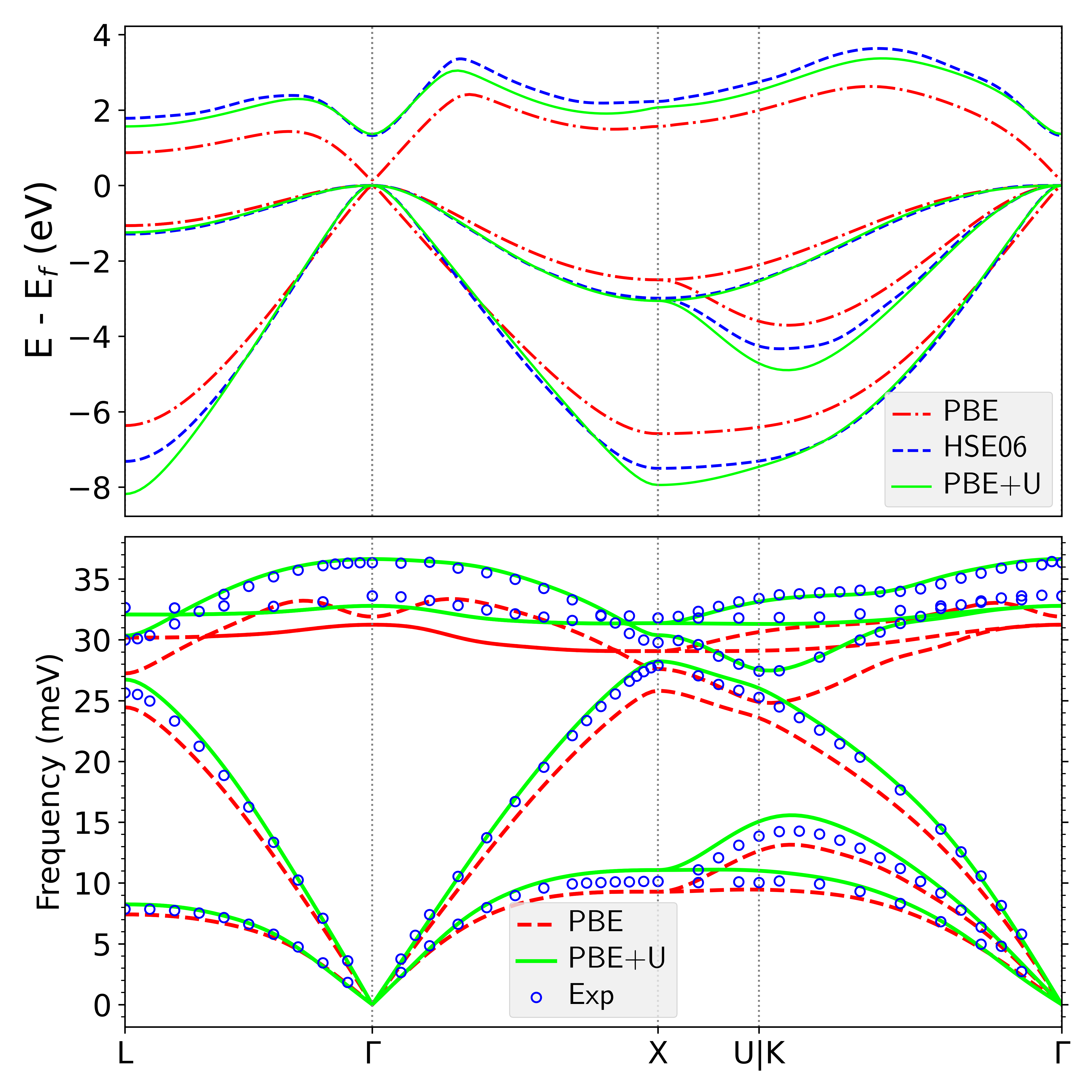}
\caption{\label{fig:band_and_phonon_disp} GaAs band structure using PBE+$U$, PBE and HSE06 (upper panel), and phonon dispersion of GaAs using PBE and PBE+$U$ comparing with experimental data from Ref \cite{Strauch_1990} (lower panel).}
\end{figure}

 These enhancements are further illustrated in Fig.~\ref{fig:band_and_phonon_disp}, where the upper panel compares the band structure obtained using PBE+$U$, HSE06, and PBE, while the lower panel depicts the phonon dispersion from PBE and PBE+$U$ alongside experimental results. The impact of the Hubbard correction on phonon dispersion is especially noteworthy, as PBE+$U$ demonstrates excellent agreement with experimental data, whereas PBE fails particularly in capturing the LO-TO splitting at the $\Gamma$ point. 

\par Overall, our PBE+$U$ results for phonon properties align well with previous computational studies~\cite{GaAs_EPW_Ma2018, Lindsay_2013}, which primarily rely on LDA. However, the accuracy of our band structure and lattice constant predictions surpasses those prior works. For instance, our electron and hole effective mass values are significantly more precise than those reported by \citeauthor{GaAs_EPW_Ma2018} using LDA. This higher accuracy is crucial for reliable transport properties calculations, as will be discussed in the following sections. 

\subsection{\label{subsec:WF}Wannier-Interpolated Band Structure and Its Impact on Mobility}
We use the \textsc{Perturbo} code~\cite{Perturbo_code_2021} to calculate phonon-limited electron mobility by solving LBTE using ITA based on the WF technique. The \textsc{Wannier90} code~\cite{W90_Pizzi, W90_Arash} is commonly employed to generate WFs in the form of maximally localized Wannier functions (MLWF). A key challenge in constructing MLWFs is the selection of projection functions used to construct the initial guess for the unitary transformation. Several approaches exist for this purpose. The first approach is the use of traditional hydrogenic orbitals ($s$, $p$, $d$ or hybrids like $sp^3$), which often requires manual tweaking even for relatively simple systems. The second utilizes pseudo-atomic orbitals (PAO), which can be directly derived from the used pseudopotential~\cite{PAO_Agapito_2016, PDWF_Qiao2023} and have proven efficient for many systems. The third option is the selected columns of density matrix (SCDM) method~\cite{scdm_Damle2017, scdm_Vitale2020}, which provides a more automated and versatile workflow. However, achieving accurate band interpolation with SCDM often requires a denser $\mathbf{k}$-mesh~\cite{PDWF_Qiao2023}.

\par Another crucial factor is the size of the coarse $\mathbf{k}$-mesh used for the WF interpolation, which directly influences the quality of the MLWF band structures. Well-converged MLWFs enable precise interpolation of e-ph matrix elements, ensuring reliable transport property calculations. One widely used metric for evaluating the accuracy is the band distance \cite{PDWF_Qiao2023, scdm_Vitale2020}, defined as the root mean squared error between the band structures obtained from the MLWFs and from DFT, with a Fermi-Dirac weighting scheme applied to downweight errors at energies far from the Fermi level. Wannierization, as a tight-binding approximation, typically provides more accurate representations of valence bands than conduction bands, and its interpolation accuracy for higher conduction bands progressively decreases with energy. Since this work focuses on electron mobility in GaAs, primarily involving the solution of the BTE for the lowest conduction band, the band distance evaluation can be restricted to this band. While band distance provides a direct measure of interpolation accuracy, an alternative and more practical approach is to compare the effective mass extracted from the two band structures, as this quantity reflects the band curvature near the CBM, which is a key factor influencing mobility. As will be shown in the following, this consideration is particularly important in high-mobility polar materials like GaAs, where small deviations in effective mass can lead to significant variations in mobility. Prioritizing effective mass over band distance is also computationally efficient, as it requires fewer energy points for evaluation. Thus, verifying the convergence of effective mass with respect to the coarse $\mathbf{k}$-mesh size serves as a robust criterion for achieving accurate mobility predictions.

\par To shed some light on the importance of a well-converged effective mass in Wannier-interpolated bands, we compute the room-temperature electron drift mobility of GaAs as a function of the coarse $\mathbf{k}$-mesh size used to set up the MLWFs. These calculations are based on the DFT+$U$ band structure results described in subsection \ref{subsec:band-structure}. MLWFs are constructed using four valence and four conduction bands. Three Wannierization techniques are tested: (i) the hydrogenic method employing hybrid $sp^3$ orbitals centered on Ga and As atoms; (ii) the SCDM method with parameters being automatically determined using the protocol proposed in Ref. \cite{scdm_Vitale2020}; and (iii) the PAO method, considering only the $s$ and $p$ projections of Ga and As atoms derived automatically from their pseudopotentials.

\par Next, the band distance within \SI{1}{\electronvolt} above the CBM and the electron effective mass at the CBM are computed for the first conduction band. The LBTE was solved using ITA, following the methodological details explained in Ref. \cite{Perturbo_code_2021}. Since in these calculations the size of the coarse $\mathbf{k}$ and $\mathbf{q}$ grids must be commensurate, and in order to analyze the convergence trends in smaller steps for the sake of clarity, DFPT+$U$ calculations were performed using a suboptimal $4\times4\times4$ $\mathbf{q}$-point mesh. The LBTE was solved for energies up to \SI{250}{\milli\electronvolt} above the CBM and the e-ph matrix elements were interpolated on dense $400\times 400\times 400$ $\mathbf{k}$ and $\mathbf{q}$ meshes. We ensured through convergence test that these parameters produce stable mobility values. The carrier concentration was set to \SI{1e12}{\per\cubic\centi\meter} to characterize the intrinsic mobility.  

\par
\begin{figure*}[bthp]
\centering
\includegraphics[scale=0.65]{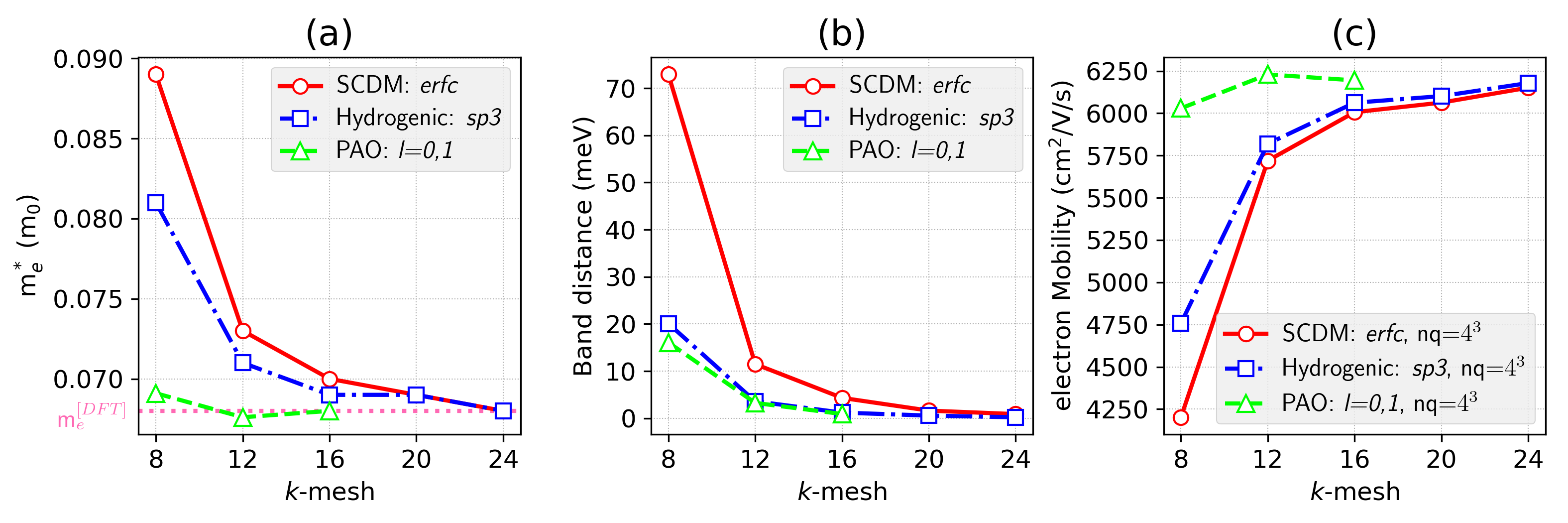}
\caption{\label{fig:me_and_mu_vs_k} Effect of the coarse $\mathbf{k}$-grid size on (a) the effective mass, (b) the band distance for the first Wannier-interpolated conduction band, and (c) the calculated electron mobility for different Wannierization methods.}
\end{figure*}

 Fig.~\ref{fig:me_and_mu_vs_k} summarizes the results of these calculations. 
As expected, increasing the coarse $\mathbf{k}$-mesh size improves the accuracy of the Wannier-interpolated band and leads to the convergence of the mobility value. Among the three Wannierization techniques, the PAO method exhibits the fastest convergence. Even at a small $\mathbf{k}$-mesh size of 8, PAO yields a sufficiently accurate effective mass of 0.069, while increasing the mesh size by one step already reproduces the DFT value of $m_e=$~\SI{0.068}{m_0}. Further increase in $\mathbf{k}$-mesh size does not change the effective mass but only slightly reduces the band distance to below 1 meV, resulting in mobility changes of less than 1\%. In contrast, the Hydrogenic and SCDM methods converge much slower, requiring a $\mathbf{k}$-mesh size of at least 16 to ensure a reliable mobility value. Notably, SCDM exhibits particularly poor convergence, though its automated workflow still makes it more convenient to use in complex systems compared to the hydrogenic method. The importance of prioritizing the effective mass over the band distance for convergence is evident. For instance, at a mesh size of 8, PAO and the hydrogenic method yield nearly similar band distances, yet their effective masses differ significantly, leading to substantial variations in mobility. 

\par We also report the use of the recently developed projectability-disentangled Wannier functions technique, which has been reported to yield more accurate Wannier-interpolated bands with smaller $\mathbf{k}$-mesh sizes when combined with PAOs \cite{PDWF_Qiao2023}, now available from \cite{qiao_github_repo}. However, since this method was not included in the latest official release of \textsc{Wannier90} and was not yet supported for transport calculations in \textsc{Perturbo} at the time of our study, we merely limit our discussion to noting that our preliminary results confirm improved convergence of the effective mass and band distance compared to only using the PAO method. 

\subsection{\label{subsec:trans} Effective Mass-dependent Drift and Hall Electron Mobilities}
The unsatisfactory overestimation of GaAs electron mobility using ITA based on LDA calculations~\cite{Perturbo_code_2021} has been addressed in several studies. \citeauthor{GaAs_GW_EPW_Liu2017} demonstrated that using the GW method to refine the band structure leads to more accurate results, where ITA yields electron mobility in agreement with experimental data, while SERTA underestimates it. \citeauthor{GaAs_EPW_Ma2018} later confirmed these findings, reporting an effective mass of \SI{0.061}{m_0} and a $\Gamma$-L valley energy of \SI{0.28}{\electronvolt} for GW calculations. \citeauthor{perturbo_GaAs_electron-two-phonon_Lee2020} derived a computationally expensive electron-two-phonon (2ph) scattering process and applied it alongside the conventional electron-one-phonon (1ph) mechanism in LDA-based simulations. Their results indicate that this additional scattering channel reduces mobility values, causing SERTA to underestimate mobility, while ITA results become more consistent with experimental data. However, since the effective mass remained underestimated in their calculation, they further attempted to refine the effective mass via manually rescaling LDA eigenvalues by a factor of $m_e^{\mathrm{exp}}/m_e^{\mathrm{LDA}}=0.067/0.055$, resulting in too low mobility values compared to experimental data \cite{perturbo_GaAs_electron-two-phonon_Lee2020_supp}. 

\par Compared to previous studies, our computational setup is based on the PBE+$U$ method, which as is clear from Table~\ref{tab:table1}, reproduces a more accurate lattice parameter and band structure. Nevertheless, a more important characteristic of this method is that it provides an efficient solution to analyze the impact of band features, particularly the effective mass, on electron transport properties. This is achieved by varying the Hubbard parameters to reproduce bands with different electron effective mass values. As will be shown, this approach not only allows for investigating the relationship between effective mass and mobility but also provides an efficient means to account for temperature-induced changes in the band structure, thereby refining temperature-dependent electron mobility predictions. This approach is a more constructive solution than manually rescaling eigenvalues by a factor~\cite{perturbo_GaAs_electron-two-phonon_Lee2020_supp} or using the parabolic band approximation~\cite{epw_ponce_Hall_mobility_2021}, since in this way other characteristics of the conduction band, such as nonparabolicity, are preserved and also the e-ph matrix is recalculated consistently with the band structure.

\par We begin our analysis by detailing the selection of Hubbard parameters corresponding to distinct effective mass values. The experimental data presented in Table. \ref{tab:table1} for $m_e$ and $\Delta E_{L-\Gamma}^{c}$ correspond to low-temperature values. A review of the available experimental studies on temperature dependence of these parameters suggests that both decrease with increasing temperature \cite{GaAs_review_Blakemore}. Specifically, for $\Delta E_{L-\Gamma}^{c}$, an analytical relation proposed in Ref. \cite{GaAs_review_Blakemore} based on the work of \citeauthor{cband_minima_aspnes} predicts a monotonically decreasing, nonlinear trend from its low-temperature value to a range of \SIrange{0.29}{0.27}{\electronvolt} between \SI{150}{\kelvin} and \SI{500}{\kelvin}. Similarly, for $m_e$, a decrease with temperature has been confirmed by \citeauthor{room_temp_me_Stradling1970}, reporting a value of \SI{0.0636}{m_0} at \SI{290}{\kelvin}. To explore the impact of effective mass variation, we adjusted the Ga and As Hubbard parameters to generate multiple band structures with different $m_e$ values. The lattice constant was kept fixed at the value used in the DFPT+$U$ calculations. Given that the $\Gamma$-L valley energy reported in Table~\ref{tab:table1} is underestimated, we constrained our $U$ parameter search to maintain it at \SI{0.28}{\electronvolt}, which aligns better with experimental data and accurately matches the GW results from Ref.~\cite{GaAs_EPW_Ma2018}. This ensures that intervalley scattering processes are not artificially introduced by an incorrectly scaled L valley energy relative to the conduction band minimum at the $\Gamma$ point.

\par
\begin{figure}[htbp]
\centering
\includegraphics[scale=0.6]{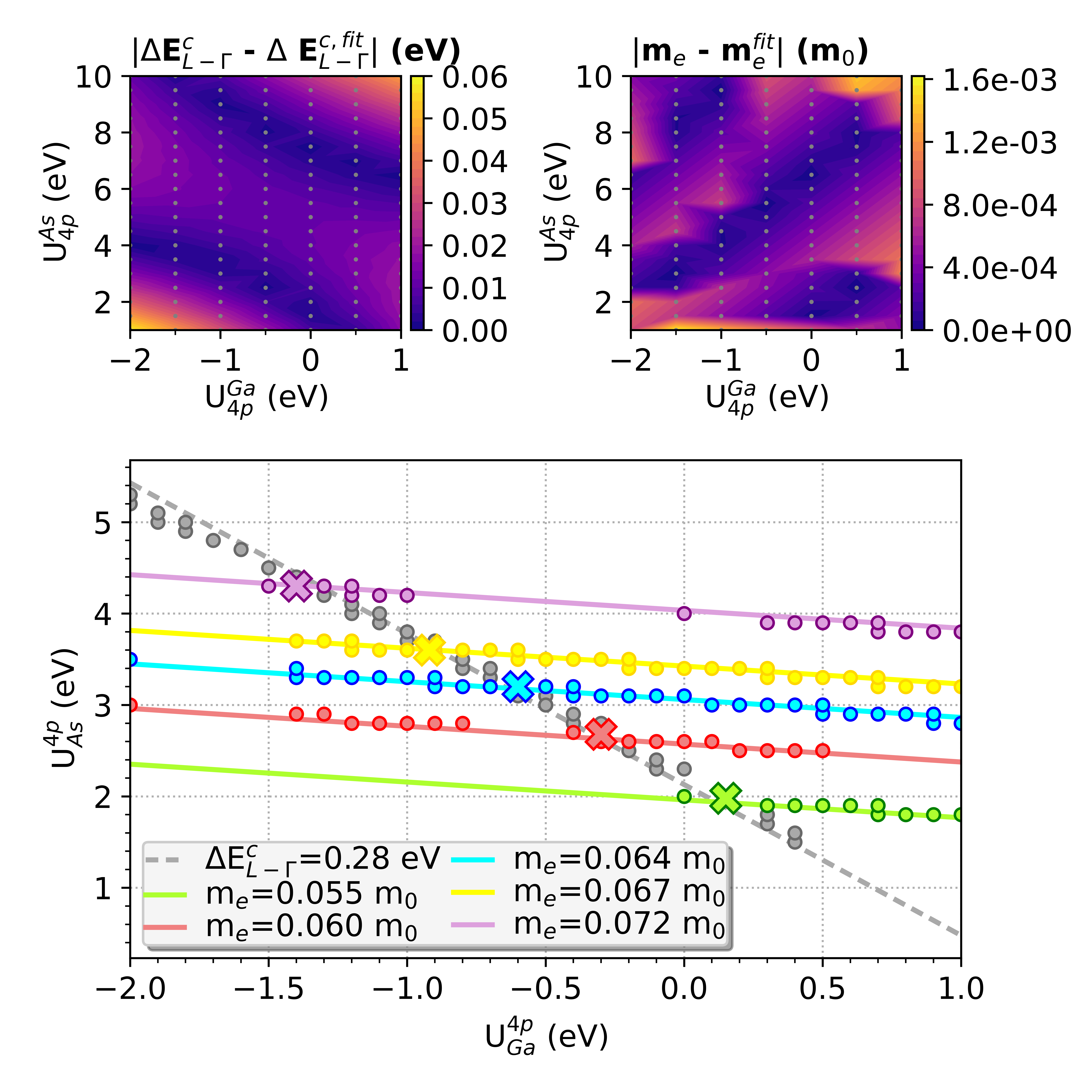}
\caption{\label{fig:vary_u}Illustration of how linear fitting aids in selecting Hubbard parameter sets corresponding to various effective masses, while keeping the $\Gamma$–L valley fixed. The upper panels show the absolute fitting errors for the $\Gamma$–L valley (left) and effective mass (right) with respect to the actual DFT data across the $U$ space. The lower panel presents the linear fitting functions corresponding to the chosen effective masses (solid lines), and the fixed $\Gamma$–L valley (dashed line). The DFT data used for the fitting are indicated by round markers, while the final selected Hubbard parameter sets for each effective mass are marked with bold crosses.}
\end{figure}

 In principle, a Bayesian optimization procedure could be employed to determine the optimal set of $U$ values for each effective mass. However, our analysis of the variation of band characteristics as a function of Hubbard parameters during Bayesian optimization revealed that each calculated quantity exhibits a nearly linear and monotonic dependence on each $U$ parameter, albeit with different slopes and small fluctuations. Based on this behavior, we perform a linear fit expressed as
\begin{equation}
    \Xi^\mathrm{fit} = \alpha U_{4p}^{Ga} + \beta  U_{4p}^{As} + \Xi_0\,,
    \label{eq:U_fit}
\end{equation}
where $\Xi$ represents the target band characteristic. The upper panels of Fig.~\ref{fig:vary_u} demonstrate the accuracy of this fitting approach for both the $\Gamma$–L valley energy and the electron effective mass by plotting the absolute error between the fitted values and those obtained from DFT calculations. As is evident by the small errors, the linear fitting method provides a reliable approximation. 

\par The fitted function can be used to find the point in the $U$-space which corresponds to the desired value of $\Xi$. Since we have two independent $U$ parameters, the solution space for a specific band feature forms a line in the $U$-space. Therefore, to find the $U$ pair that simultaneously yields target values for both the effective mass and the $\Gamma$–L valley energy, the intersection point of the corresponding lines is identified, and a focused search is performed in its vicinity, either using Bayesian optimization or a manually refined mesh, until the target values are met. This process is shown for the finally selected $U$ sets in the lower panel of Fig.~\ref{fig:vary_u}. Each optimized final $U$ set is close to the intersection point of the lines of constant $m_e$ and the $\Delta E^c_{L-\Gamma}=$\SI{0.28}{\electronvolt}  line,  indicating the robustness of this method.  

\par
\begin{figure}[htbp]
\centering
\includegraphics[scale=0.5]{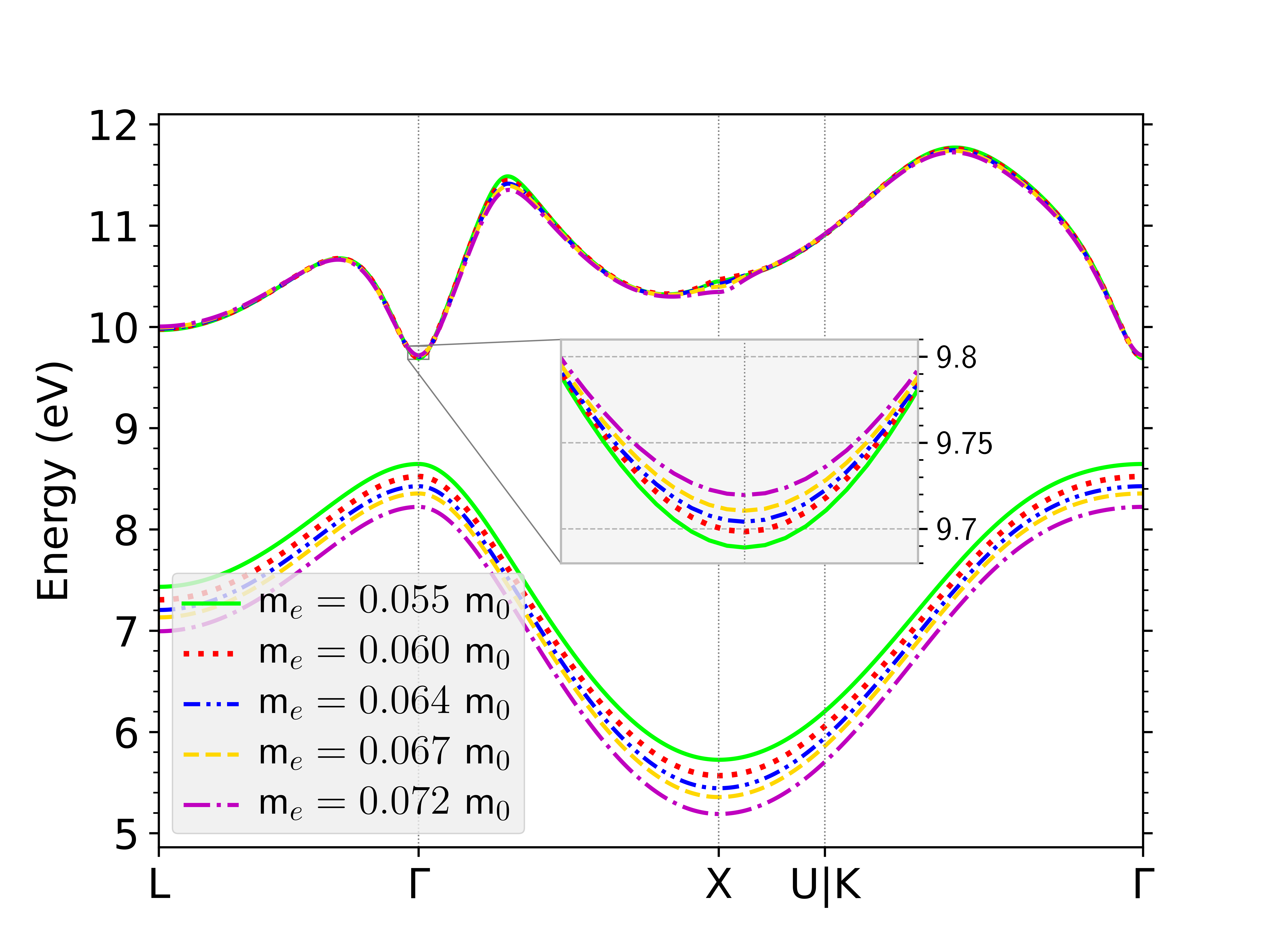}
\caption{\label{fig:band_of_diff_me} Valence and conduction bands of GaAs for various set of Hubbard parameters corresponding to different electron effective masses given in Table \ref{tab:table2}. The zoomed inset displays the conduction band near the $\Gamma$ point.}
\end{figure}

 Table \ref{tab:table2} presents the results of this process for selected effective mass values. In all cases, the constraint on the $\Gamma$-L valley energy was successfully maintained. Additionally, all parameter sets resulted in a non-zero band gap, which, while not directly influencing transport calculations, remains an important physical requirement. To ensure the physical validity of the generated band structures, we carefully examined the band structure for all chosen Hubbard parameter sets, confirming that the variations introduced do not lead to any nonphysical effects. For further illustration, Fig.~\ref{fig:band_of_diff_me} depicts these band structures near the Fermi level. The zoomed-in inset highlights the conduction band curvature at the $\Gamma$ point, emphasizing the systematic variation of the effective mass.

\begin{table}[htbp]
\caption{\label{tab:table2} Optimized Hubbard parameters for different electron effective masses while $\Gamma$-L valley is kept constant. Hubbard parameters and band edges are given in \unit{\electronvolt}, while the effective mass is given as multiple of \unit{m_0}. Here, the lattice constant is fixed at the value reported in Table \ref{tab:table1}.} 
\begin{ruledtabular}
\begin{tabular}{llcccc}
\textrm{$U_{Ga}^{4p}$} &\textrm{$U_{As}^{4p}$} &\textrm{$\Delta E_g$} &\textrm{$\Delta E_{L-\Gamma}^{c}$} &\textrm{$\Delta E_{X-\Gamma}^{c}$} &\textrm{$m_e$}\\[0.2em]
\colrule
 0.15  &1.98  &1.04 &0.28 &0.76 &0.055\\
-0.30  &2.68  &1.18 &0.28 &0.77 &0.060\\
-0.60  &3.20  &1.28 &0.28 &0.72 &0.064\\
-0.92  &3.60  &1.36 &0.28 &0.69 &0.067\\
-1.40  &4.30  &1.49 &0.28 &0.62 &0.072\\
\end{tabular}
\end{ruledtabular}
\end{table}

\par The obtained band structures corresponding to each effective mass were used to calculate the temperature-dependent electron mobility through solving the LBTE using ITA. For these calculations, we use the DFPT+$U$ setup specified in Sec.~\ref{subsec:band-structure}, in particular an $8\times8\times8$ $\mathbf{q}$-mesh was used. Consequently, MLWFs were constructed using PAOs and a $16\times16\times16$ coarse $\mathbf{k}$-mesh, ensuring convergence of the effective mass and sufficiently small band distance for the Wannier-interpolated bands. Corrections for the dynamical dipole interactions in polar materials corrections~\cite{GaAs_SERTA_Zhou2016} as implemented in the \textsc{Perturbo} package were incorporated in all calculations. The carrier concentration was kept fixed at \SI{1e12}{\per\cubic\centi\meter}, characterizing intrinsic mobility condition. The $\mathbf{k}$ and $\mathbf{q}$ meshes used to interpolate the e-ph matrix elements were determined through convergence tests on the mobility, where convergence was considered achieved when variations remained below 1\%. Accordingly, $400\times400\times400$ $\mathbf{k}$ and $\mathbf{q}$ grids were used for interpolation. Initial scattering rates for the ITA were computed within the Fan-Migdal formalism using 5 million $\mathbf{q}$-points randomly sampled from a Cauchy distribution \cite{GaAs_SERTA_Zhou2016}. The LBTE was solved for energies up to \SI{350}{\milli\electronvolt} above the CBM. We verified that this energy range yields converged mobility values at the highest temperature considered in our study, \SI{500}{\kelvin}.

\par Before discussing the results for the macroscopic mobility, it is useful to first investigate the individual e-ph scattering rates. Reasonably accurate e-ph scattering rates have been reported by previous \textit{ab initio} studies based on LDA~\cite{GaAs_SERTA_Zhou2016, GaAs_EPW_Ma2018} and GW~\cite{GaAs_GW_EPW_Liu2017, GaAs_EPW_Ma2018}. These works calculate the Fan-Migdal self-energy $\Sigma_{n\mathbf{k}}$ from which the scattering rate follows as $\tau^{-1}_{n\mathbf{k}}=2\mathrm{Im}\,\Sigma_{n\mathbf{k}}/\hbar$. This method allows for a mode-by-mode analysis of the e-ph scattering contributions from different phonon branches. However, since our focus here is solely on ITA-based results, we compute the effective scattering rates, $\tau^{-1}_{n\mathbf{k}, \mathrm{eff}}$, and the scalar effective mean free path (MFP), $\Lambda_{n\mathbf{k}}$, after converging the iterative solver following the approach introduced in Ref. \cite{GaAs_GW_EPW_Liu2017}, using
\begin{equation}
\begin{split}
&\tau^{-1}_{n\mathbf{k}, \mathrm{eff}} = \left(\frac{\mathbf{F}_{n\mathbf{k}} \cdot \mathbf{v}_{n\mathbf{k}}}{|\mathbf{v}_{n\mathbf{k}}|^2}\right)^{-1} \\
&\Lambda_{n\mathbf{k}} = \frac{\tau^{-1}_{n\mathbf{k}, \mathrm{eff}}}{|\mathbf{v}_{n\mathbf{k}}|}
\end{split}
\label{eq:eff_sca}
\end{equation}
where $\mathbf{F}_{n\mathbf{k}}$ represents the converged mean-free displacement obtained after ITA convergence. 

\begin{figure}[htbp]
\centering
\includegraphics[scale=0.5]{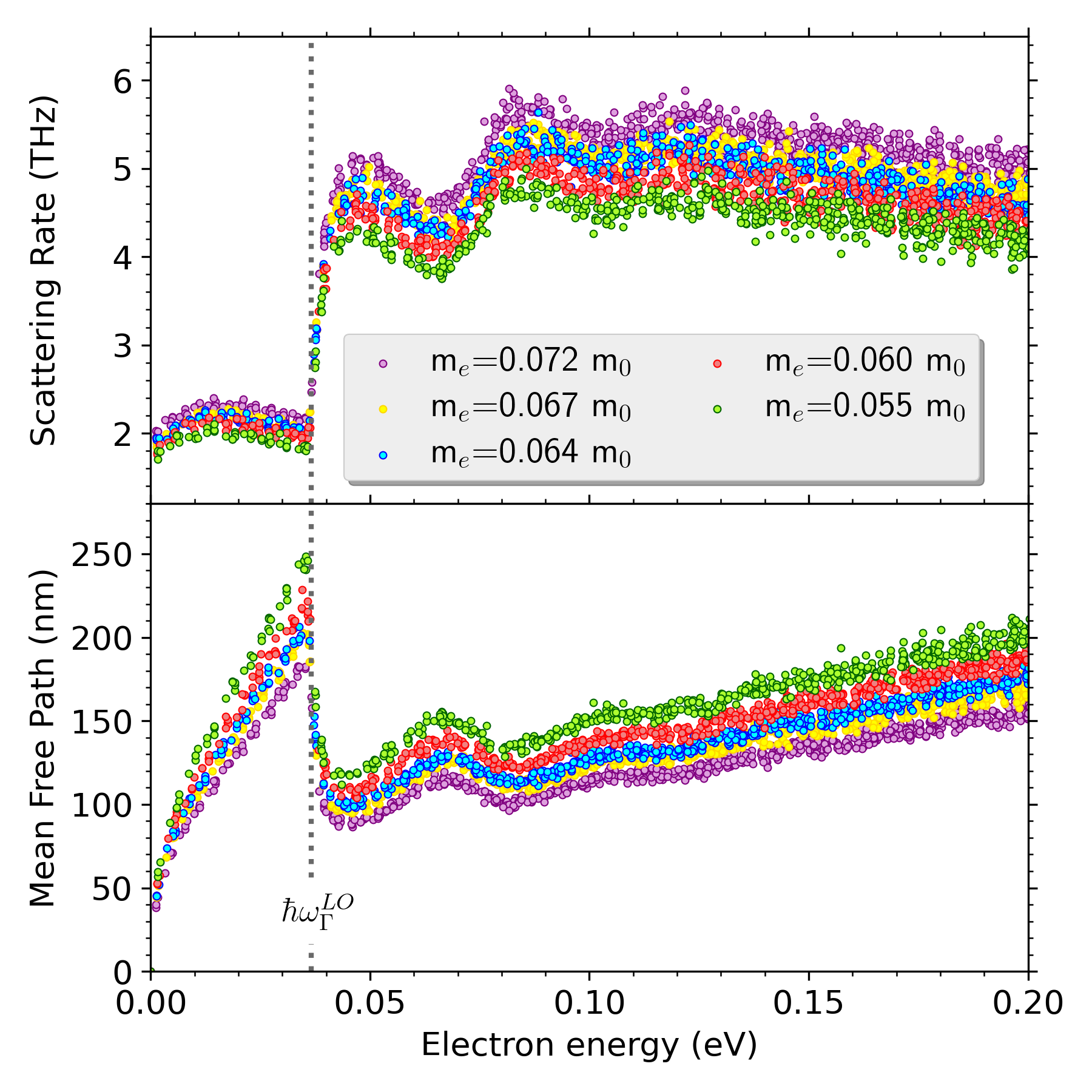}
\caption{\label{fig:scatt_rates_ITA}Room-temperature effective e-ph scattering rates (upper panel) and the corresponding effective mean free path (lower panel) as a function of electron energy above the CBM calculated using Eq. \ref{eq:eff_sca} for different effective masses.}
\end{figure}

\par Fig.~\ref{fig:scatt_rates_ITA} presents the results of these calculations at room temperature. The effective scattering rates (upper panel) show good agreement with experimental measurements reported in Ref. \cite{exp_scat_rates_Levi1985}, while the mean free path (MFP) results (lower panel) are consistent with the experimentally observed range of \SIrange{126}{130}{\nano\meter} at \SI{85}{\milli\electronvolt} above the conduction band minimum (CBM), as reported in Ref. \cite{exp_mfp_scat_Heiblum1989}. Comparisons with previous studies based on LDA \cite{GaAs_SERTA_Zhou2016} and GW \cite{GaAs_GW_EPW_Liu2017, GaAs_EPW_Ma2018} reveal several common trends. Notably, a sharp increase in the scattering rate (or equivalently, a drop in MFP) occurs at approximately \SI{0.035}{\electronvolt}, which has been attributed to strong electron–longitudinal optical (LO) phonon interactions \cite{GaAs_SERTA_Zhou2016, GaAs_GW_EPW_Liu2017, GaAs_EPW_Ma2018}. Moreover, as clearly shown in the results, scattering rates increase (or MFP decreases) with increasing effective mass, in agreement with analytical models which show an explicit $\tau^{-1} \propto \sqrt{m_e}$ dependency~\cite{Lundstrom2000, GaAs_GW_EPW_Liu2017}.

\par
\begin{figure*}[htbp]
\centering
\includegraphics[scale=0.54]{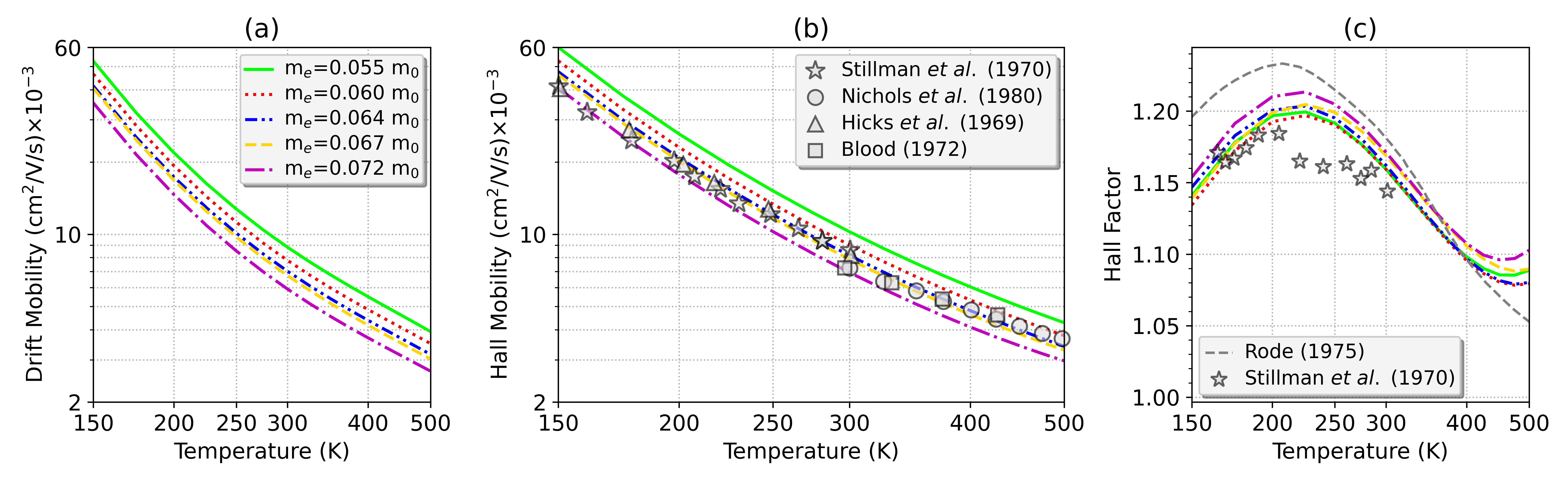}
\caption{\label{fig:ITA_mobility_vs_temp_differnt_me}Temperature-dependent electron drift mobility (a), Hall mobility (b), and Hall factor (c) calculated for selected sets of Hubbard parameters corresponding to various effective masses from Table \ref{tab:table2}. The experimental data for the Hall mobility are taken from Refs. \cite{expe_mobility_Stillman1970, expe_mobility_Nichols1980, expe_mobility_Hicks1969, expe_mobility_Blood1972}. In the Hall factor graph, experimental data represented by star markers are taken from Ref. \cite{expe_mobility_Stillman1970}, and theoretical data based on semi-empirical models represented by gray dashed lines are taken from Ref. \cite{Rode1975}.}
\end{figure*}

 Fig.~\ref{fig:ITA_mobility_vs_temp_differnt_me}(a) illustrates the temperature-dependent drift mobility results for the selected effective masses. We compare our room-temperature mobility values with previously reported \textit{ab initio} ITA-based results. Compared to the results of \citeauthor{perturbo_GaAs_electron-two-phonon_Lee2020} obtained using LDA, our mobilities for $m_e=$~\SI{0.055}{\unit{m_0}} and $m_e=$~\SI{0.067}{\unit{m_0}} are more consistent with their reported values that include the additional 2ph scattering mechanism. For instance, for $m_e=$~\SI{0.055}{\unit{m_0}}, our calculated mobility is \SI{8832}{\square\centi\meter\per\volt\per\second}, whereas their reported value accounting for 1ph+2ph scattering is approximately \SI{8600}{\square\centi\meter\per\volt\per\second}, compared to about \SI{15500}{\square\centi\meter\per\volt\per\second} when the conventional 1ph mechanism is considered. Furthermore, our result for $m_e=$~\SI{0.060}{\unit{m_0}} is \SI{7783}{\square\centi\meter\per\volt\per\second}, which is comparable to reported values in GW-based studies: \SI{8340}{\square\centi\meter\per\volt\per\second} and \SI{8985}{\square\centi\meter\per\volt\per\second}~\cite{GaAs_EPW_Ma2018} corresponding to $m_e=$~\SI{0.061}{\unit{m_0}}. It is important to note, however, that precise comparisons may be unreliable due to differences in computational details. For example, none of the previous studies have verified the effective mass convergence of Wannier-interpolated bands, which as shown in the previous subsection is a key factor influencing the final mobility values.

\par Most previous works compare their calculated \textit{ab initio} drift mobilities with available experimental Hall mobility data, assuming that the Hall factor for this material is approximately unity. Hall mobility can be calculated by solving BTE in the presence of small magnetic field. An applicable methodology is introduced in two independent studies~\cite{magnetotransport_pertubo, epw_ponce_Hall_mobility_2021} and applied to investigate Hall mobilities and Hall factors for several semiconductors, including GaAs. However, in both od these studies, the results for GaAs electron Hall mobility using ITA overestimate experimental data. In the case of Ref.~\cite{epw_ponce_Hall_mobility_2021}, this overestimation can be attributed to their crude approximation of the conduction band as parabolic with an effective mass of $m_e=$~\SI{0.067}{\unit{m_0}}, thereby neglecting the effects of band nonparabolicity. In Ref.~\cite{magnetotransport_pertubo}, the overestimation arises from the underestimation of the electron effective mass due to using LDA. Following their methodology, we have calculated the Hall electron mobility and the Hall factor using a small magnetic field of \SI{1e-8}{\tesla} for the selected effective masses. The temperature-dependent results for these quantities is compared to available experimental data in panels (b) and (c) of Fig.~\ref{fig:ITA_mobility_vs_temp_differnt_me}.

\par We begin by discussing our Hall mobility results. At lower temperature, our data corresponding to $m_e=$~\SI{0.067}{\unit{m_0}} show the best agreement with experimental data. The discrepancy between the various experimental datasets at room temperature is most likely due to the different experimental setup, such as the applied magnetic field intensity. Considering the temperature dependency of the effective mass which yields a value of $m_e=$~\SI{0.064}{\unit{m_0}} at room temperature~\cite{room_temp_me_Stradling1970, GaAs_review_Blakemore}, our computed Hall electron mobility is \SI{8161}{\square\centi\meter\per\volt\per\second}, agreeing well with experimental values ranging from \SIrange{7230}{8600}{\square\centi\meter\per\volt\per\second}. As the temperature increases, the agreement progressively shifts toward data with lower effective mass values, which is consistent with experimental observations indicating a decrease in effective mass with increasing temperature~\cite{GaAs_review_Blakemore}. For instance, our data for $m_e=$~\SI{0.060}{\unit{m_0}} increasingly aligns with experimental values, consistent with predictions of a $k \cdot p$ models that estimate $m_e=$~\SI{0.060}{\unit{m_0}} at 500 K \cite{GaAs_review_Blakemore}. 

\par The results for the Hall factor, shown in Fig.~\ref{fig:ITA_mobility_vs_temp_differnt_me}(c),
are compared with theoretical predictions from Ref. \cite{Rode1975}, obtained by solving the BTE using semi-empirical models, and with experimental data from Ref. \cite{expe_mobility_Stillman1970}, measured at a magnetic field of 0.05 T. As illustrated, all simulations corresponding to different effective masses yield nearly identical Hall factor values, indicating no significant dependence on the effective mass. Importantly, our results consistently predict values greater than 1 across the temperature range considered, which is expected due to polar optical phonon scattering being the dominant mechanism in this regime \cite{Rode1975}. Notably, excellent agreement with experimental data is observed at temperatures below 200 K, while at higher temperatures, up to room temperature, our results show only minor discrepancies and still exhibit better consistency with experiment than the semi-empirical predictions.

\par
\begin{figure*}[htbp]
\centering
\includegraphics[scale=0.58]{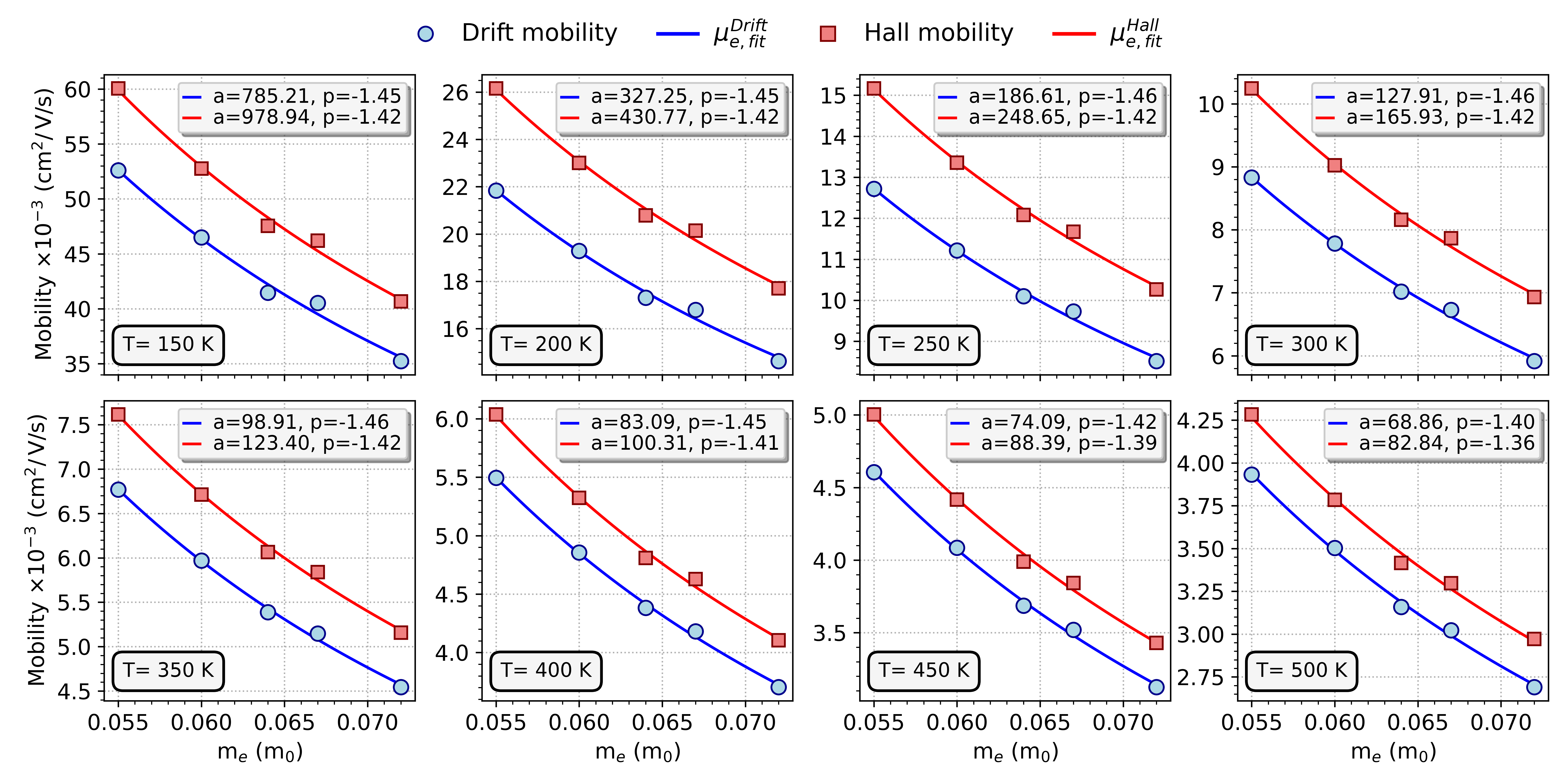}
\caption{\label{fig:power_law_fit}Fitting of computed drift and Hall electron mobility results as a function of electron effective mass using a power-law relation, $\mu_{e, \mathrm{fit}}=a\,m_e^p$, at different temperatures. In each panel, solid lines represent the fitted functions for drift and Hall mobilities, while markers indicate the computed data for the selected effective masses. The fitted coefficients $a$ and exponents $p$ are provided in the legend of each panel.}
\end{figure*}

 To investigate the relationship between the calculated mobilities and the electron effective mass, it is instructive to consider semi-empirical relations commonly used to estimate electron mobility in polar materials. These models typically predict an explicit power-law dependence of mobility on effective mass, expressed as $\mu \propto m_e^{-3/2}$ \cite{analytical_relation_polar_mobility_Gelmont1995}. In light of this, we fit our calculated drift and Hall mobility results at each temperature using a general power-law function of the form $\mu = a m_e^p$. Figure \ref{fig:power_law_fit} presents the results of these fits for both types of mobility at several selected temperatures.

\par As shown in Fig.~\ref{fig:power_law_fit}, our computed data adhere closely to the expected power-law relation. In particular, the fitted power exponent $p$ at each temperature, provided in the legend of each panel, is approximately \num{-1.45} for drift mobility and \num{-1.42} for Hall mobility, which are in close agreement with the explicit power \num{-1.5} derived in semi-empirical models. The slight discrepancy stems from the contribution of the band nonparabolicity, which is inherently considered accurately in our calculations and is only approximated in semi-empirical relation \cite{analytical_relation_polar_mobility_Gelmont1995} by analytical integrations with band-gap and temperature dependencies (See Eq. 2 of Ref. \cite{analytical_relation_polar_mobility_Gelmont1995}). Since the influence of the nonparabolicity term becomes more significant at higher temperatures, it leads to a modest reduction in the magnitude of the fitted exponents above 400 K. This trend can also be attributed to the weakening dominance of polar optical phonon scattering at elevated temperatures, as reflected in the Hall factor results (Fig.~\ref{fig:ITA_mobility_vs_temp_differnt_me}), where the Hall factor decreases and approaches 1. This consistency between \textit{ab initio} calculations and long-established semi-empirical relations highlights the robustness of our approach.

\par To further validate the reliability of our fitting, we tested the fitted functions against mobility values computed for additional effective masses. The results showed excellent agreement. For example, using the GGA-PBE functional without Hubbard correction and employing the lattice parameter reported in Table~\ref{tab:table1}, the calculated band structure yields a band gap of $\Delta E_g =$~\SI{0.65}{\electronvolt} and an effective mass of $m_e =$~\SI{0.037}{\unit{m_0}}. At room temperature, this setup produces an electron drift mobility of \SI{15633}{\square\centi\meter\per\volt\per\second}, while the fitted power-law function predicts \SI{15758}{\square\centi\meter\per\volt\per\second}, demonstrating high accuracy over a wide range of effective masses. 

\section{\label{sec:conc}conclusions}
\par In this work, we have investigated the influence of band structure features, particularly the electron effective mass, on the temperature-dependent drift and Hall mobilities of GaAs using \textit{ab initio} simulations. These simulations are based on DFT and DFPT, combined with the solution of the LBTE via the iterative approach, using the WF technique. Electronic structure and phonon calculations were carried out using PBE+$U$ and DFPT+$U$ methods, incorporating Hubbard corrections on the Ga and As 4$p$ orbitals. In this context, it is shown that applying a Bayesian optimization procedure to tune the Hubbard parameters against a reference HSE06 dataset yields an optimized band structure with accurately refined key features, including the band gap, electron and hole effective masses, and inter-valley energy separations, all in close agreement with experimental data. Furthermore, phonon properties calculated using DFPT+U method, based on this optimized band structure, show excellent agreement with experimental values in phonon dispersions and dielectric constants.

\par Given the critical influence of e-ph interactions near the CBM on electron mobility in polar material, we demonstrate that the electron effective mass of the Wannier-interpolated conduction band serves as a more robust and physically meaningful metric for assessing Wannierization accuracy than the commonly used band distance metric. In this respect, our results show that achieving convergence in the effective mass of the Wannier-interpolated bands ensures the convergence of computed mobility values. Furthermore, among the different projection schemes tested, the pseudo-atomic orbital (PAO) method outperforms the hydrogenic and SCDM approaches by delivering accurate and converged effective masses with coarser $\mathbf{k}$-mesh sizes, making it a more efficient choice for high-throughput or large-scale transport calculations.

\par Building on previous studies that highlight the impact of band structure refinement, particularly effective mass, on electron mobility of GaAs, we have shown that the DFT+$U$ method can be effectively employed to systematically explore the relationship between electron mobility and effective mass. This is achieved by varying the Hubbard parameters to generate band structures with different effective mass values, while keeping the $\Gamma$–L inter-valley energy separation fixed. This approach inherently preserves the nonparabolicity of the conduction band, a key characteristic that significantly influences the final mobility values. Moreover, it allows for a self-consistent calculation of the e-ph matrix elements corresponding to each modified band structure. The nearly linear dependence of key band features on the Hubbard parameters greatly simplifies the search for optimal $U$ values corresponding to target band features by allowing the search space to be effectively constrained around an estimated parameter set.

\par The outcome of the aforementioned considerations is reflected in our transport results obtained using the iterative solution of the LBTE. First and foremost, our calculated e–ph scattering rates, Hall mobility, and Hall factor exhibit excellent agreement with experimental data. In particular, our calculated Hall mobility at a given temperature and a specific effective mass is in excellent agreement with the temperature-dependent effective mass extracted from measurements and $k\cdot p$ models. This demonstrate the effectiveness of our method in accounting for temperature-dependent band structure effects on transport properties of materials, either by extracting them from experimental data or, as is recently being pursued, using direct \textit{ab initio} simulations \cite{temperature-dependent_band_ponce}. 

\par Moreover, the calculated drift mobilities are comparable to other reported \textit{ab initio} results based on GW, emphasizing the critical role of accurate band structure refinement, particularly the electron effective mass, in determining mobility values using the iterative solution of the BTE. While GW calculations usually impose expensive computational cost, specifically for complex systems, our results show that the DFT+U approach provides a more computationally efficient and flexible alternative, as it allows targeted tuning of band features, making it well suited for systematically studying their impact on transport properties.

Last but not least, investigating the relationship between mobility and effective mass by fitting our calculated data to a power-law relation of the form $\mu \propto m_e^p$ reveals a strong agreement with established theory. Specifically, the fitted exponent for drift mobility, $p \approx -1.45$, closely matches the well-known semi-empirical value of $-1.5$. The slight deviation is attributed to the nonparabolic nature of the conduction band, which introduces an implicit dependence on effective mass, an effect that semi-empirical models can only capture approximately. 

\section{\label{sec:ack}Acknowledgments}
This research has received funding from the Austrian Research Promotion Agency FFG (project number 895289). The financial support by the Austrian Federal Ministry for Digital and Economic Affairs, the National Foundation for Research, Technology and Development and the Christian Doppler Research Association is gratefully acknowledged. Calculations were performed using supercomputer resources provided by the Vienna Scientific Cluster (VSC).

\section{Data Availability}
The data that support the findings of this study are available from the corresponding author upon
reasonable request.

\section{Conflict of Interest}
The authors have no conflicts to disclose.

\bibliography{refs}

\end{document}